\documentclass[twocolumn]{aastex62}

\usepackage{xcolor}

\defcitealias{2015ApJ...805...35H}{HL+15}

\graphicspath{{./}{figures/}}

\submitjournal{ApJL}

\shorttitle{Extreme Feedback in SPT-CLJ0528-5300}
\shortauthors{Calzadilla et al.}

\begin{document}

\title{Discovery of a Powerful ${>}10^{61}$ erg AGN outburst in the Distant Galaxy Cluster SPT-CLJ0528-5300}

\correspondingauthor{Michael Calzadilla}
\email{msc92@mit.edu}

\author[0000-0002-2238-2105]{Michael S. Calzadilla}
\affil{Kavli Institute for Astrophysics and Space Research, Massachusetts Institute of Technology Cambridge, MA 02139, USA}

\author[0000-0001-5226-8349]{Michael McDonald}
\affiliation{Kavli Institute for Astrophysics and Space Research, Massachusetts Institute of Technology Cambridge, MA 02139, USA}

\author[0000-0003-1074-4807]{Matthew Bayliss}
\affiliation{Department of Physics, University of Cincinnati, Cincinnati, OH 45221, USA}
\affiliation{Kavli Institute for Astrophysics and Space Research, Massachusetts Institute of Technology Cambridge, MA 02139, USA}

\author[0000-0002-5108-6823]{Bradford A. Benson}
\affil{Kavli Institute for Cosmological Physics, University of Chicago, 5640 South Ellis Avenue, Chicago, IL 60637, USA}
\affil{Fermi National Accelerator Laboratory, MS209, P.O. Box 500, Batavia, IL 60510, USA}
\affil{Department of Astronomy and Astrophysics, University of Chicago, 5640 South Ellis Avenue, Chicago, IL 60637, USA}

\author[0000-0001-7665-5079]{Lindsey E. Bleem}
\affil{Kavli Institute for Cosmological Physics, University of Chicago, 5640 South Ellis Avenue, Chicago, IL 60637, USA}
\affil{High Energy Physics Division, Argonne National Laboratory, 9700 S. Cass Avenue, Argonne, IL 60439, USA}

\author[0000-0002-4208-798X]{Mark Brodwin}
\affil{Department of Physics and Astronomy, University of Missouri, 5110 Rockhill Road, Kansas City, MO 64110, USA}

\author{Alastair C. Edge}
\affil{Centre for Extragalactic Astronomy, Department of Physics, Durham University, Durham, DH1 3LE, UK}

\author[0000-0003-4175-571X]{Benjamin Floyd}
\affil{Department of Physics and Astronomy, University of Missouri, 5110 Rockhill Road, Kansas City, MO 64110, USA}


\author{Nikhel Gupta}
\affil{School of Physics, University of Melbourne, Parkville, VIC 3010, Australia}

\author[0000-0001-7271-7340]{Julie Hlavacek-Larrondo}
\affil{D\'{e}partement de Physique, Universit\'{e} de Montr\'{e}al, Montr\'{e}al, Qu\'{e}bec H3C 3J7, Canada}

\author[0000-0002-2622-2627]{Brian R. McNamara}
\affil{Department of Physics and Astronomy, University of Waterloo, Waterloo, ON N2L 3G1, Canada}
\affil{Waterloo Centre for Astrophysics, University of Waterloo, Waterloo, ON N2L 3G1, Canada}
\affil{Perimeter Institute for Theoretical Physics, 31 Caroline Street North, Waterloo, ON N2L 2Y5, Canada}

\author[0000-0003-2226-9169]{Christian L. Reichardt}
\affil{School of Physics, University of Melbourne, Parkville, VIC 3010, Australia}

\collaboration{(SPT collaboration)}




\begin{abstract}

We present ${\sim}$103 ks of \textit{Chandra} observations of the galaxy cluster SPT-CLJ0528-5300 (SPT0528, $z{=}0.768$). This cluster harbors the most radio-loud ($L_{\rm 1.4 GHz} {=}~ 1.01{\times}10^{33}$ erg s$^{-1}$ Hz$^{-1}$) central AGN of any cluster in the South Pole Telescope (SPT) SZ survey with available X-ray data. We find evidence of AGN-inflated cavities in the X-ray emission, which are consistent with the orientation of the jet direction revealed by ATCA radio data. The combined probability that two such depressions --  each at ${\sim}1.4-1.8\sigma$ significance, oriented ${\sim}180^{\circ}$ apart and aligned with the jet axis -- would occur by chance is 0.1\%. At ${\gtrsim}10^{61}$ erg, the outburst in SPT0528 is among the most energetic known in the universe, and certainly the most powerful known at $z{>}$0.25. This work demonstrates that such powerful outbursts can be detected even in shallow X-ray exposures out to relatively high redshifts ($z {\sim} 0.8$), providing an avenue for studying the evolution of extreme AGN feedback. The ratio of the cavity power ($P_{\rm cav} {=}~ 9.4 {\pm} 5.8 {\times}10^{45}$ erg s$^{-1}$) to the cooling luminosity ($L_{\rm cool} {=}~ 1.5 {\pm} 0.5 {\times} 10^{44}$ erg s$^{-1}$) for SPT0528 is among the highest measured to date. If, in the future, additional systems are discovered at similar redshifts with equally high $P_{\rm cav}/L_{\rm cool}$ ratios, it would imply that the feedback/cooling cycle was not as gentle at high redshifts as in the low-redshift universe.

\end{abstract}



\section{Introduction}\label{sec:intro}

Early X-ray studies of the intracluster medium (ICM) in galaxy clusters revealed central cooling times that were often much less than the age of the Universe \citep[e.g.][]{1977ApJ...215..723C,1977MNRAS.180..479F,1991MNRAS.252...72W,1992MNRAS.258..177E}. Theory predicted that ``cooling flows'' should result in massive reservoirs of cold gas deposited onto the central galaxies, along with high star formation rates \citep[see review by][]{1994ARA&A..32..277F}. However, multi-wavelength investigations into these ``downstream'' observables of cooling flows found them to be an order of magnitude lower than predicted \citep[e.g.][]{1987MNRAS.224...75J,1989AJ.....98.2018M,1995MNRAS.276..947A,1999MNRAS.306..857C,2006ApJ...652..216R,2008ApJ...681.1035O,2015ApJ...805..177D,2018ApJ...858...45M}. A proposed solution to this long-standing issue came in the form of feedback from active galactic nuclei (AGN), where the energy output from an accreting supermassive black hole (SMBH) in the central galaxy prevents excessive cooling of the ICM out of the hot, X-ray emitting phase \citep[][]{1993MNRAS.264L..25B,2000A&A...356..788C,2001ApJ...554..261C,2007ARA&A..45..117M,2012NJPh...14e5023M,2012ARA&A..50..455F}. This heating by the AGN, which is probed by measuring the sizes of bubbles inflated in the hot ICM by radio jets, has been found to correlate strongly with several cooling properties of the ICM, implying a tightly-regulated feedback loop \citep[e.g.][]{2004ApJ...607..800B,2004MNRAS.355..862D,2006ApJ...652..216R,2010ApJ...720.1066C,2011ApJ...740...51M,2012MNRAS.421.1360H,2013ApJ...777..163H,2013ApJ...774...23M,2017MNRAS.464.4360M,2017MNRAS.471.1766B}. 

Recently, surveys taking advantage of the redshift-independent Sunyaev-Zel'dovich (SZ) effect, such as by the South Pole Telescope \citep[SPT;][]{2011PASP..123..568C}, have enabled studies of the \emph{evolution} of AGN feedback over cosmic time, revealing no significant evolution in the cooling properties of clusters \citep{2013ApJ...774...23M,2017ApJ...843...28M} or in the heating properties of AGN  \citep[e.g.][]{2012MNRAS.421.1360H,2015ApJ...805...35H} out to $z{\sim}1.2$. In particular, \citet{2015ApJ...805...35H} studied AGN feedback in SPT clusters in a redshift range $0.4\leq z \leq 1.2$, and find that the mechanical feedback by AGN in brightest cluster galaxies (BCGs) has, on average, remained unchanged for over half the age of the Universe. However, \citet{2017MNRAS.471.1766B} find some evidence of evolution in the radio-luminosity function of SZ-selected clusters \citep[see also][]{2017MNRAS.464.4360M,2019arXiv190611388G}, showing that high luminosity radio sources have a higher occurrence rate at higher redshifts. This could indicate a transition from high-excitation radio galaxy accretion modes to low-excitation accretion modes at intermediate redshifts which is possibly driven by enhanced galaxy merger rates at higher redshifts that trigger AGN activity \citep[e.g.][]{2013ApJ...773..154L,2013ApJ...779..138B}. Thus, understanding the extreme outbursts often associated with high radio luminosity is crucial for understanding the co-evolution of radio sources with their host galaxies and cluster environments.

Here we investigate the effects of AGN feedback in the SZ-selected galaxy cluster SPT-CLJ0528-5300, at a redshift of $z{=}0.768$ and mass of $M_{500} = 3.65{\pm}0.73 {\times} 10^{14}$ $M_{\odot}$ \citep{2015ApJS..216...27B}. 
In \autoref{sec:obs} we summarize the data used in this paper. In \autoref{sec:results} we report our detection of large X-ray cavities and argue for their credibility in \autoref{sec:discuss}. We discuss the implications of this discovery in \autoref{sec:implications} before summarizing our results in \autoref{sec:end}. We assume a $\Lambda$CDM cosmology with $H_0 {=} 70$ km s$^{-1}$ Mpc$^{-1}$, $\Omega_m {=} 0.27$ and $\Omega_\Lambda {=} 0.73$. All errors are 1$\sigma$ unless noted otherwise.

\begin{figure*}
\centering
\includegraphics[width=\textwidth]{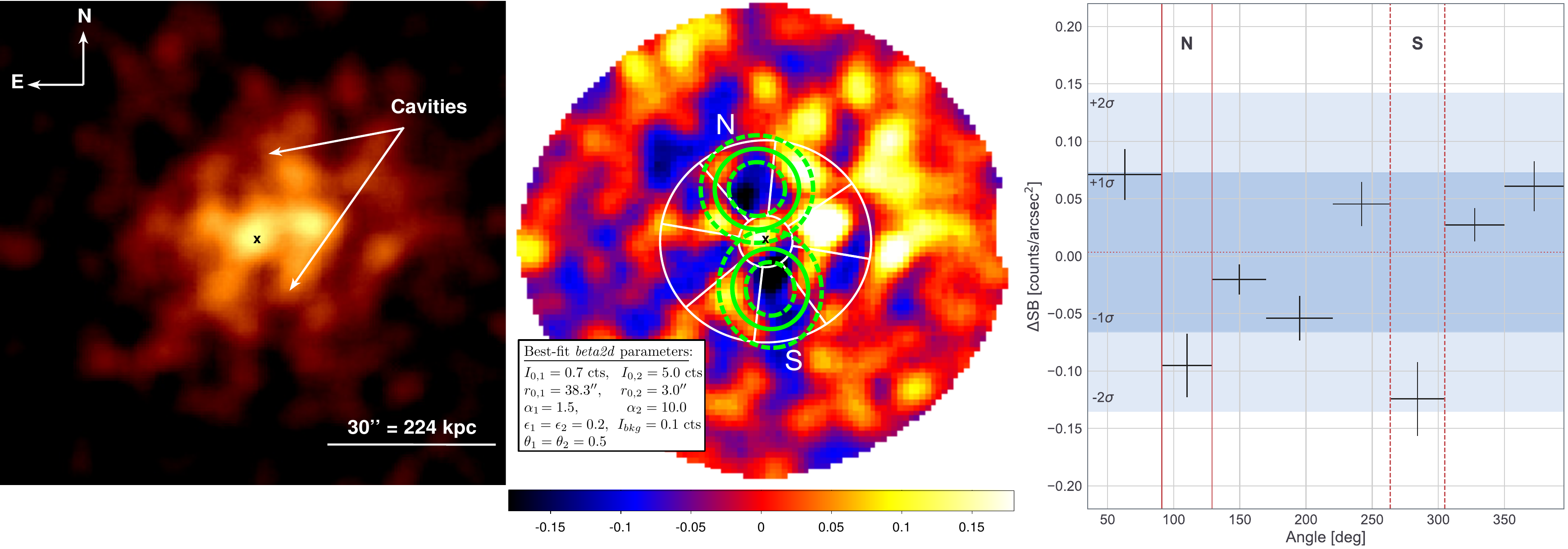}
\caption{\textbf{Left:} stacked, cleaned 0.5--4 keV \textit{Chandra} counts image of SPT0528, and \textbf{center:} double-beta model-subtracted residual image with best-fit parameters, both on the same spatial scale. The `x' marks the location of the large-scale ICM centroid. The counts image has been smoothed with a 10-pixel radius Gaussian kernel, while the residual image was first binned by $2{\times}2$-pixels then smoothed with a 5-pixel Gaussian kernel. Solid and dashed ellipsoidal regions outline the detected cavities and their estimated size uncertainty. White annular wedges illustrate where we measured surface brightness in the \emph{unsmoothed} residual map as a function of azimuthal angle to determine the significance of the cavities, shown in the \textbf{right} panel. The two cavities are detected at 1.4$\sigma$ and 1.8$\sigma$ below the median.
\label{fig:xray}}
\end{figure*}

\section{Cluster Selection \& Data Analysis} \label{sec:obs}

The galaxy cluster SPT-CLJ0528-5300 (hereafter, SPT0528) was selected via the SZ effect as part of the 2500 deg$^2$ SPT-SZ survey \citep{2015ApJS..216...27B}. The SZ effect is a particularly useful mechanism enabling the detection of distant galaxy clusters independent of redshift \citep[e.g.][]{2009ApJ...701...32S,2013ApJ...763..127R,2013JCAP...07..008H,2014A&A...571A..29P}. The SPT-SZ catalog\footnote{\url{https://pole.uchicago.edu/public/data/sptsz-clusters/}} found 516 clusters, with a median sample redshift of 0.55
\citep{2015ApJS..216...27B}.

We search the fiducial SPT-SZ catalog for galaxy clusters hosting radio-loud sources, whose energy outputs are predominantly in the form of outflows that do mechanical work on their environments. To this end, we cross-match the SPT-SZ catalog with the Sydney University Molonglo Sky Survey (SUMSS) source catalog \citep{2003MNRAS.342.1117M}, which imaged 8100 deg$^2$ of the radio sky below $\delta < -30^{\circ}$ at 843 MHz. 
We find a total of 112 out of 677 clusters 
with an associated SUMSS radio source within $38\arcsec$. Comparing the source densities of the two catalogs, this is an excess over random by a factor of ${\sim}17$. The observed 0.8 GHz flux of each source is k-corrected and converted to a 1.4 GHz rest-frame radio power assuming $L \propto \nu^{\alpha}$, 
where 
$\alpha = -0.7$ is a typical power-law spectral index for a radio galaxy \citep{2002AJ....124..675C}.

Among the most powerful radio sources in this SPT+SUMSS sample are SPT0528, SPT-CLJ0351-4109 ($z{=}0.68$) and SPT-CLJ0449-4901 ($z{=}0.792$), all with $L_{\nu} > 10^{33}$ erg s$^{-1}$ Hz$^{-1}$. 
Of these three, SPT0528 hosts the most luminous unblended radio source with available X-ray data. We focus now on SPT0528, leaving the other two radio-bright clusters for future follow-up.

\subsection{Chandra}

Observations of SPT0528, totaling ${\sim}$124 ks of exposure time, were taken with the \textit{ACIS-I} instrument onboard \textit{Chandra} (Observation IDs: 9341, 10862, 11747, 11874, 11996, 12092, 13126) as part of an ongoing follow-up campaign \citep[e.g.,][]{2013ApJ...774...23M}.
These data were reduced and analyzed in a standard fashion similar to \citet{2011ApJ...738...48A} and \citet{2013ApJ...774...23M}, using the \textit{Chandra} Interactive Analysis of Observations {\tt (CIAO) v4.8.1} software with {\tt CALDB v4.7.0}. We applied the latest gain and charge-transfer inefficiency corrections, as well as improved background screening as the observations were taken in the \textsc{VFAINT} telemetry mode. Periods of high background were excluded, resulting in a total ``clean'' exposure of ${\sim} 103$ ks. Modeling of the global ICM properties was previously done in \citet{2013ApJ...774...23M} and we reuse their analysis pipeline here for our deeper observations.


\subsection{ATCA}
\label{subsec:atca_data}

SPT0528 was also observed with the higher resolution Australia Telescope Compact Array (ATCA) in two separate observing runs on 01/06/2015 (1--3 GHz, 63~mins) and 08/21/2016 (4.5--6.5 and 8.0--10.0 GHz, 25~mins), resulting in beams of $8\arcsec{\times}5\arcsec$, $4\farcs0 {\times} 2\farcs5$, and 3$\farcs5 {\times} 2\farcs0$ respectively.
The data were reduced with the 05/21/2015 release of the {\tt Miriad} software package \citep{1995ASPC...77..433S}. The phase calibrator J0524-5658 was used to create the radio maps, with some multi-faceting, but no self-calibration was necessary. The rms values for the images are 40, 30 and 55 $\mu$Jy at 1--3, 4.5--6.5, and 5.5--9.0 GHz, respectively.
The resulting images have a dynamic range of ${\sim} 1000$, ensuring sensitivity to extended emission.

\section{Detection of Large X-ray Cavities} \label{sec:results}

\autoref{fig:xray} shows the stacked, cleaned 0.5--4 keV \textit{Chandra} counts image of SPT0528. The large-scale ICM centroid ($05^h28^m05\fs2, -52^{\circ}59'50\farcs5$) is consistent with the BCG position \citep[$05^h28^m05\fs3, -52^{\circ}59'53\farcs5$;][]{2012ApJ...761...22S}.  The ICM ${\sim}$75 kpc (10$\arcsec$) to the north and south of the ICM centroid, outlined by the green ellipsoidal regions in \autoref{fig:xray}, is depressed relative to the surrounding emission, possibly indicating the presence of buoyantly-rising bubbles inflated by the central AGN. We fit the counts image with a double-beta model ({\tt beta2d}) with constant background using {\tt CIAO's} \textit{Sherpa} package, linking the centroid positions, ellipticities ($\epsilon$), and position angles ($\theta$) of the two components and allowing all the parameters to vary. The best-fit parameters and model-subtracted residual image are shown in \autoref{fig:xray}. The residual surface brightness of the southern and northern depressions represent 1.8$\sigma$ and 1.4$\sigma$ significant fluctuations from the expected surface brightness based on the statistics of eight similar-area azimuthal bins at a common distance from the cluster center. 

Assuming these cavities are real, it is possible to estimate the power of the AGN outburst that created them via the $pV$ work that they do in expanding by a volume $V$ against their surroundings at pressure $p$ \citep[e.g.][]{2004ApJ...607..800B,2005MNRAS.364.1343D}. We calculate the cavity power, $P_{cav}$, as follows:
\begin{equation}
P_{cav} = \frac{4 p V}{t_{\rm cav}},
\end{equation}
where $4pV = 4 (2 n_e kT)(\frac{4}{3}\pi R_{\rm min}^2 R_{\rm maj})$ is the total enthalpy of a cavity of prolate geometry, with semi-major (minor) axis $R_{maj} ~ (R_{min})$ filled with relativistic fluid, and $t_{\rm cav} {=}~ d_{\rm cav}/c_s$ is the age of the cavity, assuming a non-relativistic plasma. 
We calculate the best-fit central (r${\lesssim}50$ kpc) electron density and temperature values $n_{e,0}{=} (9.2{\pm}0.9){\times}10^{-3}$ cm$^{-3}$ and $kT_0{=} 4.2{\pm}1.5$ keV and estimate a sound speed $c_s{=} 1038{\pm}186$ km s$^{-1}$, and pressure $p = (1.2{\pm}0.5){\times}10^{-10}$ erg cm$^{-3}$ for the ICM in the vicinity of the cavities. The sizes of the cavities outlined in \autoref{fig:xray} were estimated by eye from smoothing the images at various scales, with a resulting 30--40\% size uncertainty reflecting the spread of all the estimates. The northern cavity has dimensions $R_{\rm maj} {\approx} R_{\rm min} {=}~ 60{\pm}19$ kpc ($7\farcs5{\pm}2\farcs5$), and is $80{\pm}8$ kpc ($10\farcs0{\pm}1\farcs0$) away from the ICM centroid. The southern cavity has dimensions $R_{\rm maj} {=}~ 59{\pm}19$ kpc ($7\farcs5{\pm}2\farcs5$), $R_{\rm min} {=}~ 54{\pm}18$ kpc ($6\farcs8{\pm}2\farcs4$), and is $69{\pm}8$ kpc ($8\farcs7{\pm}1\farcs0$) away from the ICM centroid. Using these highly-uncertain cavity size and distance measurements, we calculate a total cavity enthalpy of $4pV =(2.0{\pm}1.2) {\times} 10^{61}$ erg, leading to a cavity power of $P_{\rm cav}{=}~ (9.4 {\pm} 5.8 ) {\times}10^{45}$ erg s$^{-1}$. At ${\gtrsim}10^{61}$ erg, the energy associated with these cavities is among the highest measured of \emph{any} system.

\begin{figure}
\centering
\includegraphics[width=\columnwidth]{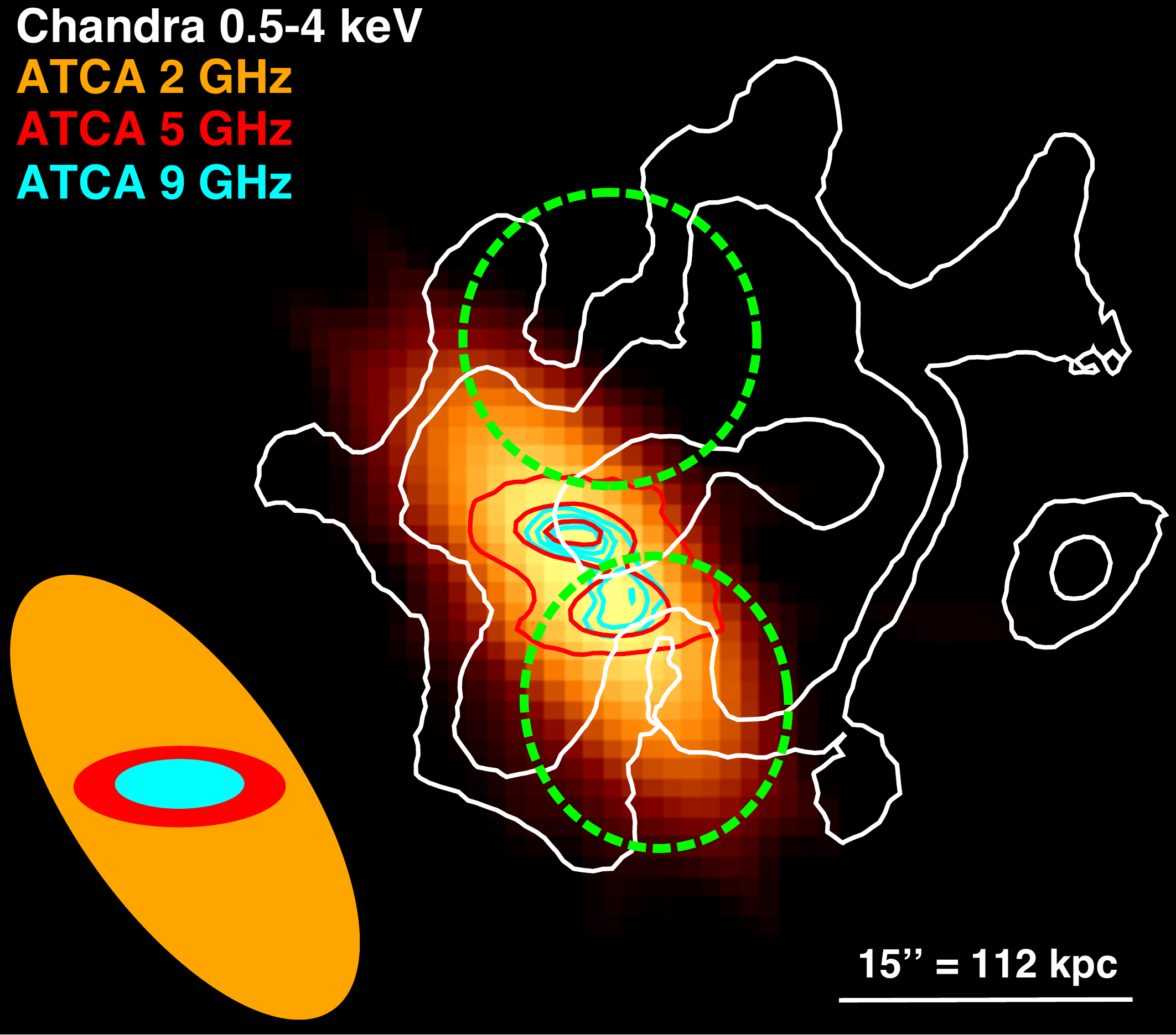}
\caption{ATCA 2 GHz radio image of SPT0528 with X-ray contours overlaid in white.
The 5 GHz ATCA data are represented by red contours at 0.4, 1.7, and 3.1 mJy beam$^{-1}$, while the 9 GHz data are represented by cyan contours at 0.5, 0.7, 1.0, and 1.3 mJy beam$^{-1}$.
The dashed green ellipsoidal regions outline the X-ray cavities as before. The radio source is coincident with the ICM centroid, and is elongated in the direction of the X-ray cavities. The ATCA beam sizes are shown in the bottom-left corner. \label{fig:atca}}
\end{figure}

\begin{figure*}
\includegraphics[width=\textwidth]{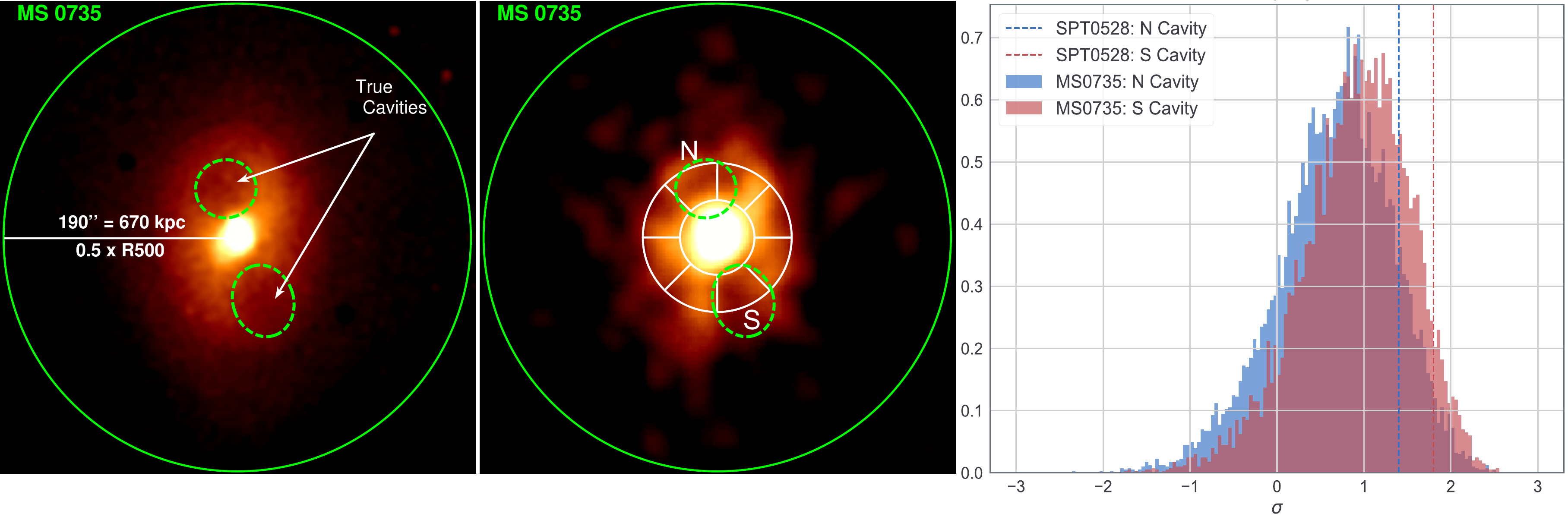}
\caption{\textbf{Left}: \textit{Chandra} image of MS0735, the most powerful AGN outburst known \citep[e.g.][]{2006ApJ...652..216R,2009ApJ...698..594M}. \textbf{Center}: MS0735 downsampled to the same depth (${\sim} 1400$ counts), energy range ($0.5{-}4$ keV), redshift ($z{=}0.768$) and self-similar scale ($r_{pixel}/R_{500}{=}~ constant$) as SPT0528. We calculate the significance of the downsampled cavities by measuring the flux in each of the eight annular wedges illustrated here, subtracting from the average flux across all eight sectors, then dividing by the scatter. \textbf{Right}: The significance distribution of the downsampled cavities in MS0735, calculated over 10,000 realizations. Dashed vertical lines indicate the significance of the cavities in SPT0528 calculated in the same way, demonstrating that the cavities in SPT0528 look similar to what a comparably powerful outburst as in MS0735 would look like at the same depth and redshift. 
\label{fig:mock_obs}}
\end{figure*}

\section{Supporting Evidence for a ${\gtrsim}10^{61}$ erg Mechanical Outburst} \label{sec:discuss}

While the statistical significance of the large cavities shown in \autoref{fig:xray} is marginal, there are other lines of evidence that support the picture of a recent ${\gtrsim}$10$^{61}$ erg outburst in the core of SPT0528 and, indirectly, increase the likelihood of these large cavities being real.

\subsection{Jet Direction \& Morphology}

The most convincing evidence in support of this cavity system comes from the simultaneous consideration of our high angular resolution X-ray and radio observations. \autoref{fig:atca} presents ATCA radio data taken at 2, 5, and 9 GHz, with \textit{Chandra} contours overlaid. The position of the radio source is consistent with that of the ICM center and BCG. At 2 GHz, the radio source is weakly resolved, while the 5 and 9 GHz contours show that the source is clearly elongated in the NE-SW direction with a possible jet component.

The position angle (PA) of the potential jet component is $205^{\circ}{\pm}15^{\circ}$, while the PA of the cavities' axes is $190^{\circ}{\pm}20^{\circ}$. These PAs are fully consistent within the uncertainties, suggesting that the X-ray emitting gas has been evacuated by the expanding radio lobes. Furthermore, the two cavities are oriented $180^{\circ}{\pm}34^{\circ}$ degrees apart with respect to the ICM centroid. To quantify our confidence in the \emph{overall} detection of these cavities, we take the product of the following independent probabilities: 
\begin{itemize}
    \item a value being ${\geq}1.4\sigma$ from the mean in only \emph{one} direction (i.e. strictly a depression): 8.1\%
    \item finding 2 such values in an 8-element array: 25\%
    \item the number of pairs in an 8-element array that are each four positions apart, divided by the total number of combinations: $8{\div}{8 \choose 2}$ = 28.6\%
    \item the jet axis aligning with any arbitrary cavity axis, given an uncertainty of ${\pm}15^{\circ}$: $30^{\circ}{\div}180^{\circ}$ = 16.7\%
\end{itemize}
Thus, the combined probability of chance alignment between the cavity axis and the jet axis, and of two $\geq1.4\sigma$ depressions separated by $\sim$180$^{\circ}$ is 0.1\%.

\subsection{X-ray Surface Brightness of the Cavities}
\label{subsec:simulations}

To investigate whether the detection significance of the cavities in SPT0528 is appropriate for a ${\gtrsim}10^{61}$ erg outburst, given the depth of these data, 
we consider a similar system with exquisitely deep data and downsample it to the same depth of our observations. Currently cited as the most energetic AGN outburst in the literature, MS0735.6+7421 (hereafter MS0735), at a redshift of $z{=}0.216$, has a total enthalpy of $4pV = 6.4 {\times} 10^{61}$ erg and has been observed for ${\gtrsim} 0.5$ Ms with \textit{Chandra} \citep{2006ApJ...652..216R,2009ApJ...698..594M}. 
We simulate what MS0735 would look like with \textit{Chandra} at the same redshift ($z{=}0.768$) and depth (${\sim}$1400 counts) as our 0.5--4 keV SPT0528 observations, by reducing the MS0735 count rate, increasing the background (noise), and resampling the image to account for the different angular diameter distance.

This downsampling procedure was repeated 10,000 times, with the results shown in \autoref{fig:mock_obs}. The full observations of MS0735 are shown in the left panel, while a single, characteristic downsampled image is shown in the middle panel. The full-depth image shows the obvious presence of cavities in the raw data, outlined by the dashed green ellipsoidal regions, which are still recovered a large fraction of the time (at an average significance of $0.9\sigma$ and $0.7\sigma$) in the significantly shallower downsampled images. In comparison, the southern and northern cavities in SPT0528 are even more convincing, at 1.8$\sigma$ and 1.4$\sigma$ below their expected surface brightness, respectively. The bootstrapping procedure above demonstrates that cavities as large as those in MS0735, the most energetic outburst we know of, could be detected with \textit{Chandra} at the same observing depth and redshift of SPT0528 at the same level, inspiring more confidence that those in SPT0528 are real. 

\begin{figure*}
\centering
\includegraphics[width=\textwidth,height=3.1in]{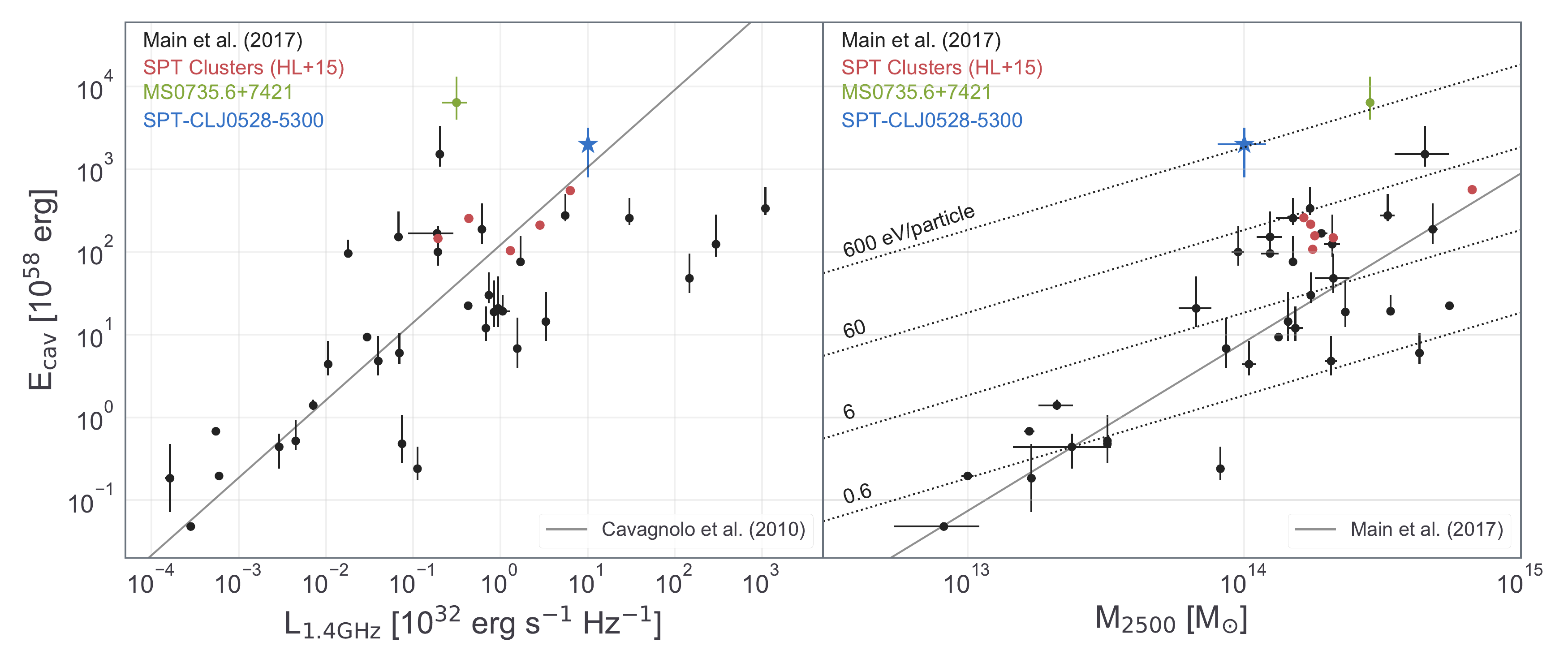}
\caption{The scaling relations between cavity/outburst energy ($E_{cav} {=}~ 4pV_{\rm tot}$) vs total radio luminosity ($L_{\rm 1.4 GHz}$) and cluster mass ($M_{2500}$), from \citet[][median $z{\sim}0.6$]{2015ApJ...805...35H} and \citet[][median $z{\sim}0.06$]{2017MNRAS.464.4360M}. SPT0528 and MS0735 are plotted here for reference. While we measure an extreme cavity energy for SPT0528, we see that it is entirely consistent with that predicted by its radio luminosity. Solid black lines in both panels represent the best-fit scaling relations among respective quantities from \citet{2010ApJ...720.1066C} and \citet{2017MNRAS.464.4360M}. \label{fig:tri_plot}}
\end{figure*}

\begin{figure*}[]
\includegraphics[width=0.5\textwidth]{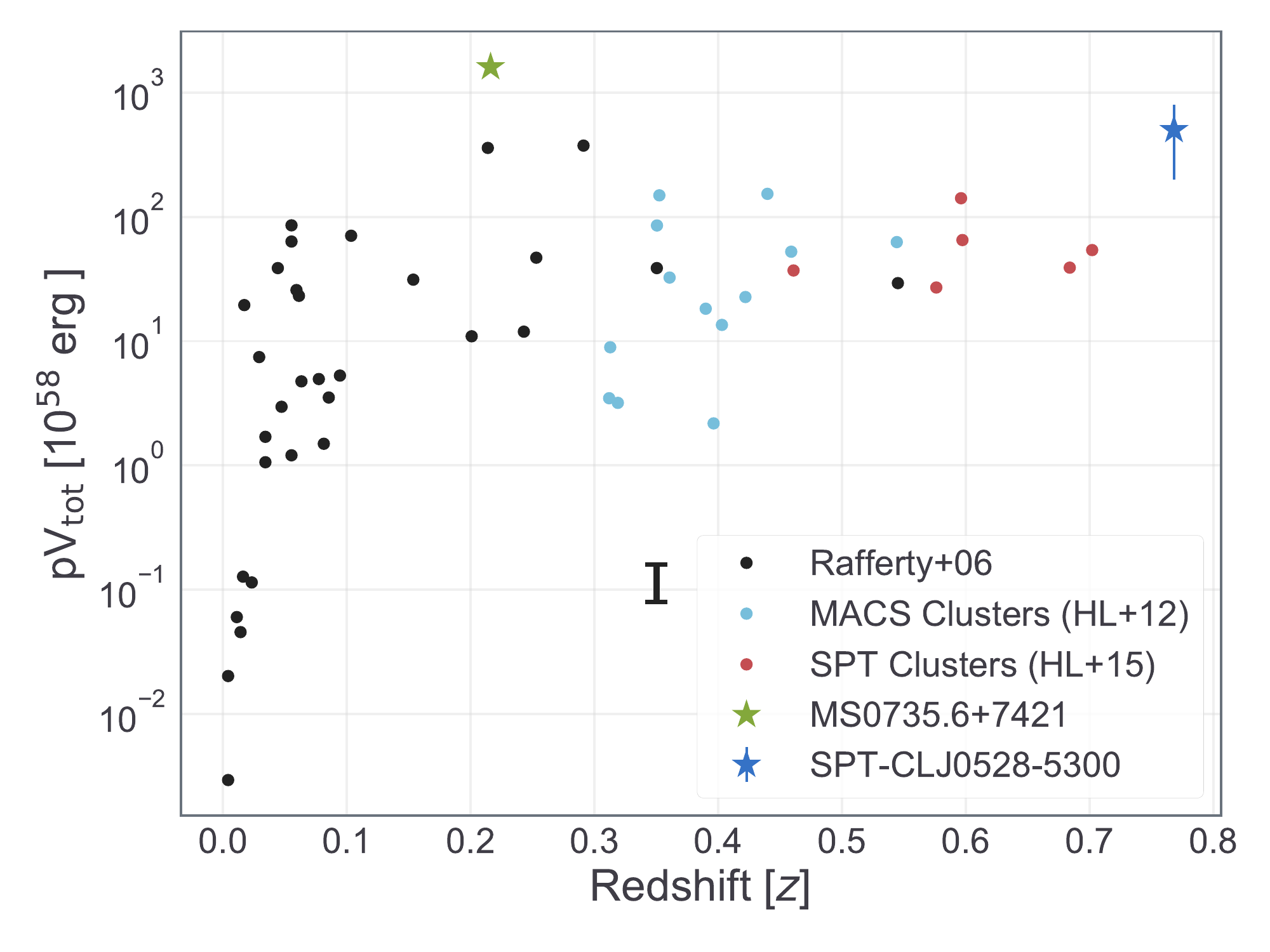}
\includegraphics[width=0.5\textwidth]{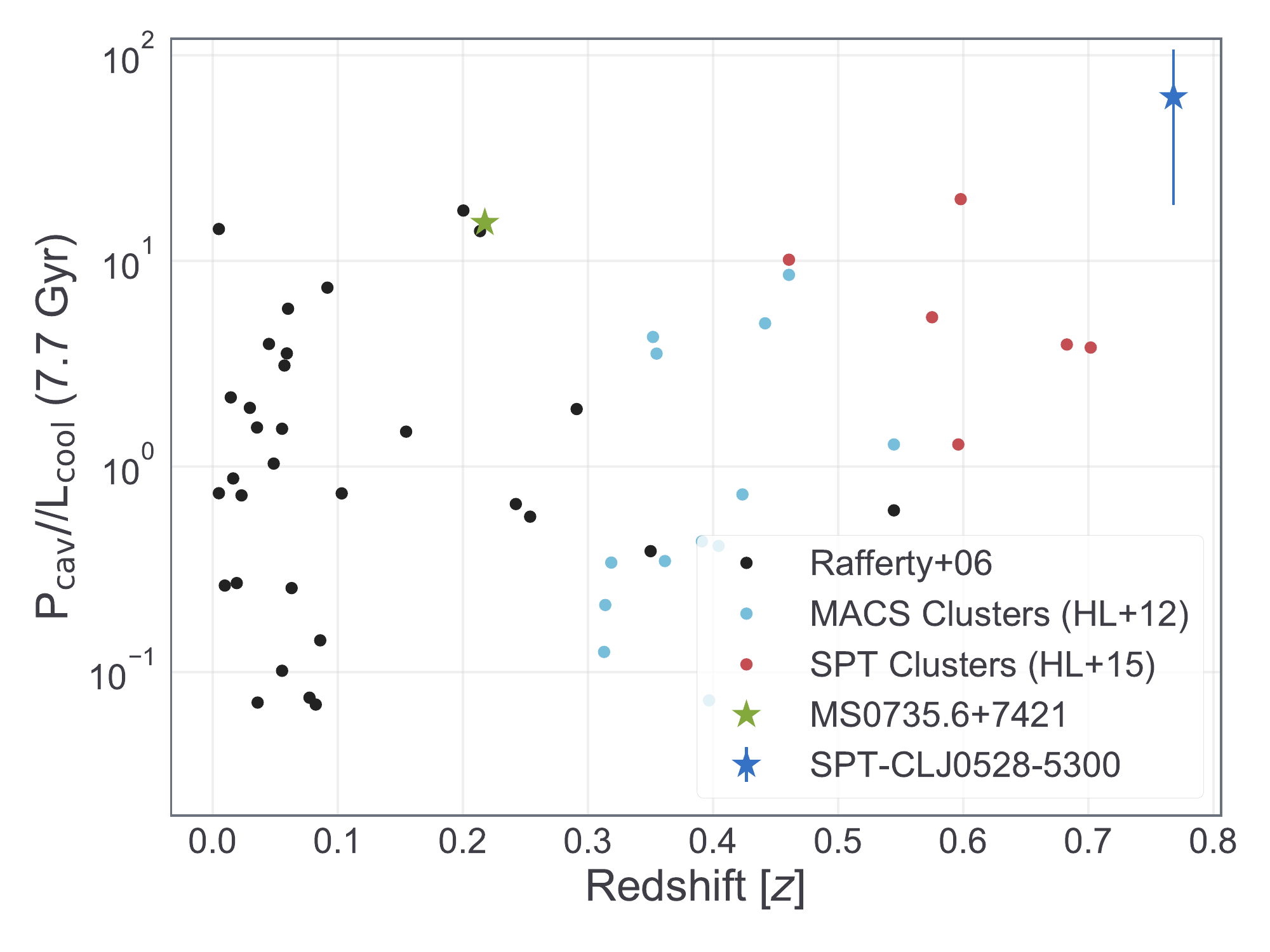}
\caption{\textbf{Left}: Cavity/outburst energy, and \textbf{right}: ratio of cavity power ($P_{\rm cav}$) to cooling luminosity ($L_{\rm cool}$), as functions of redshift, adapted from \citet{2015ApJ...805...35H}. This demonstrates that SPT0528 is amongst the most powerful outbursts yet discovered. The discovery of similar systems at equally high redshifts could significantly influence the inferred evolution of AGN feedback. \label{fig:HL15}}
\end{figure*}

\subsection{Scaling of Cavity and Radio Jet Powers}

A number of studies have established a correlation between total radio luminosity and AGN outburst powers as probed by X-ray cavities \citep[e.g.][]{2008ApJ...686..859B,2010ApJ...720.1066C,2011ApJ...735...11O}. Such a trend is to be expected as the bubbles are inflated by the radio jets. \autoref{fig:tri_plot} shows this relationship between radio power and cavity enthalpy, along with a relationship between cavity enthalpy and the host cluster mass \citep{2012MNRAS.421.1360H,2015ApJ...805...35H,2017MNRAS.464.4360M}. These relations have large scatter for high power systems, so we do not incorporate them into our overall detection probability. Nevertheless, SPT0528 was specifically chosen for follow-up as one of the most radio-loud systems in the SPT-SZ survey (see \autoref{sec:obs}), and we expect the cavity power to be correspondingly large. The extreme total cavity enthalpy of $4 p V = (2.0{\pm}1.2) {\times}10^{61}$ erg we measure is consistent with expectations given a radio luminosity of $L_{\rm 1.4 GHz}{=}~ (1.01 {\pm} 0.03) {\times} 10^{33}$ erg s$^{-1}$ Hz$^{-1}$, based on ATCA and SUMSS radio observations. Diagonal dotted lines in \autoref{fig:tri_plot} show the average energy gained per particle (assuming a gas fraction of 10\%) if the outburst energy coupled completely and isotropically to the hot gas. These lines demonstrate the similarity in energy density between the outbursts in MS0735 and SPT0528, and the significant effect this energy could have on the surrounding ICM.

\bigskip
\section{Implications for AGN Feedback at High Redshifts}
\label{sec:implications}

To determine the impact of the powerful outburst in SPT0528, we calculate a cooling luminosity, $L_{\rm cool}$. For consistency with the literature, this is the integrated luminosity within the radius where the cooling time of the ICM falls below 7.7 Gyr, or effectively, where $r\lesssim 100$ kpc \citep[e.g.][]{2008ApJ...681.1035O,2013ApJ...774...23M,2012MNRAS.421.1360H}. For SPT0528, we measure $L_{\rm cool}{=}~ (1.5 {\pm} 0.5) {\times} 10^{44}$ erg s$^{-1}$ \citep{2013ApJ...774...23M}. This measurement yields a ratio of $P_{\rm cav}/L_{\rm cool}{\approx}~ 63$, on the upper end of the typical range of other systems with cavities (see \autoref{fig:HL15}). Since $z{\sim}0.8$, $P_{\rm cav}/L_{\rm cool}$ in galaxy clusters has not shown significant evolution, implying well-regulated feedback loops. If, in the future, additional clusters exhibiting such high $P_{\rm cav}/L_{\rm cool}$ ratios are found at high redshift, it would have important implications for the inferred redshift evolution of AGN feedback.

Given that collecting sufficient X-ray counts becomes more expensive with higher redshifts, it is observationally unfeasible to systematically search for evolution in the typical mechanical powers of AGN in clusters, since we can only detect the most extreme outbursts even at modest redshift (\autoref{fig:HL15}). However, we \emph{can} search for evolution in the upper envelope of jet powers, by searching for the most extreme outbursts at each redshift. If, for example, we find that clusters at $z{\sim}1$ have significantly more AGN with cavity enthalpies $pV > 10^{61}$ ergs, it implies that feedback was more bursty than it is today. Such a conclusion would provide strong constraints for feedback models, leading to improvements in cosmological simulations.

\section{Summary} \label{sec:end}

We report the detection of a pair of extended X-ray cavities in the SZ-selected galaxy cluster SPT0528. While these cavities are marginally significant in the X-ray observations alone ($1.8\sigma$ and $1.4\sigma$), their plausibility is strengthened by additional lines of complementary evidence. First and foremost, the radio structure of this source gives us a jet direction aligned along the cavity axes, which has a 16.7\% probability of occurring by chance, making it the likely inflation mechanism. In addition, the two ${\geq}1.4\sigma$ cavities are separated by ${\sim} 180^{\circ}$ about the X-ray centroid, which should only occur randomly 0.5\% of the time. Combining these probabilities yields a 0.1\% chance that these brightness depressions are random fluctuations, equating to a Gaussian significance of ${\sim}$3.3$\sigma$. Further, SPT0528 was initially selected as being among the most radio-loud in the SPT-SZ survey, so a powerful outburst is expected, and is indeed consistent with what is predicted from scaling relations with radio power and mass. Given all of this evidence, SPT0528 appears to be an extraordinary system, and with a power of $P_{\rm cav}{=}~ 9.4 {\times}10^{45}$ erg s$^{-1}$ and total enthalpy of ${\gtrsim}10^{61}$ erg it is the most energetic $z{\gtrsim}0.25$ mechanical outburst observed yet.


\acknowledgments
We thank William Forman for helpful comments that improved the paper. Support for this work was provided to MSC and MM by NASA through Chandra Award Numbers GO7-18124 and G06-17112, issued by the Chandra X-ray Observatory Center, which is operated by the Smithsonian Astrophysical Observatory for and on behalf of the National Aeronautics Space Administration under contract NAS8-03060. Work at Argonne National Lab is supported by UChicago Argonne LLC, Operator of Argonne National Laboratory. Argonne, a U.S. Department of Energy Office of Science Laboratory, is operated under contract no. DE-AC02-06CH11357.

%

\vspace{5mm}
\facilities{CXO, SPT, ATCA}


\software{astropy, numpy, scipy, CIAO, XSPEC}

\listofchanges

\end{document}